\documentclass[aps,paper,nofootinbib,groupedaddress]{revtex4}
\usepackage[dvips]{graphicx}
\usepackage{amsmath,amssymb,graphicx,subfigure}

\newcommand{\be}{\begin{equation}}
\newcommand{\ee}{\end{equation}}
\newcommand{\bea}{\begin{eqnarray}}
\newcommand{\eea}{\end{eqnarray}}

\def\slashchar#1{\setbox0=\hbox{$#1$}           
   \dimen0=\wd0                                 
   \setbox1=\hbox{/} \dimen1=\wd1               
   \ifdim\dimen0>\dimen1                        
      \rlap{\hbox to \dimen0{\hfil/\hfil}}      
      #1                                        
   \else                                        
      \rlap{\hbox to \dimen1{\hfil$#1$\hfil}}   
      /                                         
   \fi}                                         %

\begin{document}
\preprint{ECT*-07-02}

\title{Instantons, Chiral Dynamics and Hadronic Resonances}
\author{M.~Cristoforetti, P. Faccioli and M. Traini}
\affiliation{Dipartimento di Fisica and I.N.F.N., Universit\`a degli
Studi di Trento, Via Sommarive 15, Povo (Trento) 38050 Italy.}
\affiliation{European Centre for Theoretical Studies in Nuclear
Physics and Related Areas, Strada delle Tabarelle 286, Villazzano
(Trento), I-38050 Italy.}

\begin{abstract}
We use the Interacting Instanton Liquid Model (IILM) as a tool to
study the role played by the chiral interactions in the lowest-lying
vector and axial vector meson resonances. We find that narrow $a_1$
and $\rho$ meson resonances can be generated by instanton-induced
chiral forces, even in the absence of confinement. In the IILM,
these hadrons are found to have masses only about $30\%$ larger than
the experimental value and small width $\lesssim 10-50$~MeV. This
result suggests that chiral interactions are very important  in
these systems and provide most of their mass. We explore the
decaying patterns of the $\rho$ meson, in the absence of
confinement.  We argue that, in our model where only chiral forces
are switched on, this meson decays dissociating into its quark
anti-quark constituents.
\end{abstract}

\maketitle

\section{Introduction}

Hadron spectroscopy provides basic constrains on the structure of
non-perturbative QCD dynamics. Since gluons interact with quarks in
a way which depends dramatically on the quark mass, we expect the
light- and heavy-hadron sectors of the spectrum to be sensitive to
different dynamical correlations. For example, while charmonium
spectrum is well understood in terms of linearly rising confining
potential and perturbative gluon exchange, the description of the
light-hadron spectra in terms of QCD degrees of freedom  must take
into account also the interactions responsible for spontaneous
breaking of chiral symmetry. In particular, we expect  the mass and
the structure of lowest-lying hadrons such as pion, nucleon, vector
and axial-vector mesons to be strongly influenced by the chiral
interactions, because the splitting between parity partners in this
part of the spectrum is as large as $500-600~$MeV. On the other
hand, Regge trajectories suggest that resonances with large angular
momentum are predominantly influenced by the physics of color
confinement, and it has been conjectured that chiral symmetry
breaking may be even restored up in the spectrum~(see
e.g.~\cite{glozman} and references therein). In this paper we
address the following questions: are any of the light hadron
resonances generated predominantly by chiral forces, with
confinement playing a sub-leading role? Could any of such hadrons
exist even in the complete absence of confinement? In this case,
would such systems still decay predominantly in colorless hadrons?
What is the microscopic dynamical mechanism underlying the splitting
between the different lowest-lying multiplets?

The answers to these questions root in the non-pertubative sector of
QCD. Although lattice field theory provides the only available
ab-initio tool for computing non-perturbative QCD correlation
functions, the mechanism by which the hadron structure arises is not
directly evident in such a framework. In particular, it is difficult
to disentangle the contribution to the correlation functions arising
from the different types of interaction, such as the confining
forces, the chiral forces, and the perturbative gluon exchange.
Hence, in order to gain physics insight,  in the present work we
focus on the role played by chiral symmetry breaking and we do so by
restricting the path integral to a sum over gauge field
configurations which generate the near zero-mode zone of the Dirac
operator. To this end, we adopt the interacting instanton liquid
model, developed in \cite{iilm}. Instantons have been long argued to
be the dominant fluctuations generating the zero-mode zone of the
Dirac operator, hence providing the correlations which break
spontaneously chiral symmetry. In the IILM, the QCD path integral
over all gluon configurations is replaced by an effective theory in
which the gauge fields  accounted for are those generated by
integrating over the positions, color orientations, and sizes of
singular-gauge instantons.

Several features of the instanton picture have been observed in a
number of lattice studies
\cite{latticeILM1,latticeILM2,latticeILM3,latticeILM4}. In
particular, it has been observed that chiral symmetry breaking in
QCD is strongly correlated with smooth lumps of topological gauge
fields, with a shape compatible with that of singular-gauge
instantons \cite{latticeILMlumps1,latticeILMlumps2}.

Instanton-induced correlations in hadrons have also been studied
through several phenomenological model calculations. It has been
shown that instanton models provide a  good description of the mass
and the electro-weak structure of pions, nucleons and hyperons~
\cite{instcorr,RILMm,RILMb,emff1,emff2,emff3,pion1,delta12,diquark,diakonov2}.
In a recent work ~\cite{IILMchpt}, we have used the IILM to study
the dependence of the Dirac spectrum and of the pion and nucleon
masses on the current quark mass. We have found that the dependence
of these observables on the pion mass is in quantitative agreement
with both chiral perturbation theory and lattice simulations.

The main shortcoming of the instanton liquid model is that it does
not lead to the correct behavior of the Wilson loop, i.e. it does
not predict a linearly rising potential between static color
sources, at large distances. This fact implies that, while
singular-gauge instantons accounts well for the non-perturbative QCD
correlations at the chiral symmetry breaking scale
$\Lambda_\chi\sim~1$~GeV, they do not generate sufficient long-range
correlations at the confining scale
$\sim~\Lambda_{QCD}\sim~0.2$~GeV. As a result, we expect the chiral
forces generated by these topological fluctuations to play only
marginal role in  heavy-quark systems ---which are insensitive to
chiral dynamics--- or in highly excited hadrons ---whose wave
functions are expected to extend for several fm---. On the other
hand, just because of the lack of confinement, the IILM can be used
as a tool to single-out the contribution of chiral forces  in the
different hadronic systems. By studying where the model works well,
we can identify the properties of the hadrons which are almost
completely determined by the chiral dynamics. Conversely, by
studying where the model fails, one can in principle gain
information about the structure of the additional interaction which
is needed in order to reproduce the experiment. In this context,
$a_1$ and $\rho$ resonances represent the ideal test systems: being
in between the pion --- which is strongly related to chiral symmetry
breaking--- and highly excited states --- which are dominated by
color confinement --- they are expected to be sensitive to both
chiral and confining interactions. In addition the splitting between
these two resonance masses is a direct measure of the effect of
spontaneous chiral symmetry breaking.

An exploratory study of the contribution of instanton forces to the
lowest-lying hadrons was performed in \cite{RILMm, RILMb}, where
several hadronic Euclidean point-to-point correlation functions
where calculated in the ILM up to sizes of the order of several
fractions of a fm. It was shown that instanton-induced chiral forces
are very strongly attractive in the nucleon and pion, but much
weaker in the $\rho$ and $\Delta$. Their study provides indications
that hadron resonances might exist in the instanton vacuum. In fact
$\rho$, $a_1$ and $\Delta$ masses
extracted from  a fit of the corresponding point-to-point
correlation functions using a pole-plus-continuum ansatz for the
spectral function, were found to be within $20-30~\%$ from their
experimental value.

In this model, the dynamical origin of the splitting between the
hadron  multiplets which are not connected via chiral
transformations is well understood in terms of the diluteness of the
instanton liquid. In fact, the instanton zero-mode contribution to
the two-point correlation functions associated to the pion and to
the nucleon comes at the lowest-order in the instanton liquid
diluteness $\kappa\sim~10^{-1}$. On the other hand, the analog
contribution to the correlation functions associated to vector- and
axial-vector mesons and to decuplet baryons masses comes at the
next-to-leading order, i.e. to $\mathcal{O}(\kappa^2)$. The
splitting between chiral partners is provided by the spontaneous
chiral symmetry associated with the de-localization of the fermion
zero-modes~(see \cite{rev} and references therein).

In this context an interesting question is whether, in the presence
of only instanton-induced chiral forces, the $a_1$ and $\rho$
resonances predominantly tend to decay into colorless pions, or
rather dissociate into free quarks and anti-quarks.
In the present work we address this and other related issues
concerning  the contribution of chiral dynamics to the structure of
hadronic resonances by  computing  momentum projected correlation
functions at Euclidean times up to $\sim~1.2$~fm and we focus on the
effective mass plot.

This method is usually employed in lattice QCD simulations to
extract  the masses of the stable bound-states {\it only}, i.e. when
the lowest-lying hadron contributes to the spectral density through
a $\delta$-function.

In this case, the effective mass plot displays a flat plateau in the
large Euclidean time limit.
We extend this method to include the case in which the lowest-lying
hadron is {\it not} a stable bound-state but rather a resonance with
a finite width. In this case  then effective mass plot displays an
approximatively linear and mild fall-off at intermediate Euclidean
times. This behavior can be distinguished both from the flat plateau
associated to stable bound-states and from the strong exponential
decay  associated to the filtering of a perturbative continuum. As a
result both the mass and the width of the lowest-lying mesons and
baryons can be extracted.

In addition, some information about the decay patterns of these
resonances  can be gained by studying the behavior of the effective
mass in the asymptotically large Euclidean time.  In this regime,
the effective mass converges to the smallest eigenvalue in the
transfer matrix. If the simulation box is sufficiently large,
the lowest eigenvalue of the transfer matrix in the $\rho$ and $a_1$
channels corresponds to the invariant mass of the branch-cut
singularity in their two-point function. In QCD these are expected
to be at the threshold for decaying into two- and three- pions,
respectively. In models which do not account for confinement, these
hadrons could also decay into free quarks and the branch-cut may be
shifted.

We apply the effective mass plot analysis to determine the mass of
the lowest-lying vector and axial-vector meson, using several values
of the quark mass. Our results clearly show that the
instanton-induced  chiral forces are sufficiently strong to generate
the $\rho$ meson and the $a_1$ meson resonances even in absence of
confinement.

On the other hand, while our previous studies have shown that the
nucleon and pion masses are  correctly reproduced in this
model~\cite{IILMchpt}, the vector and axial vector mesons resonances
are found to be about $30\%$ heavier than their experimental value,
with $M_\rho\simeq$~1~GeV and $M_{a_1}\simeq~1.7$~GeV.  These
results support a picture in which the mass and binding of the
nucleon and pion is almost entirely provided by chiral forces, while
the mass of the vector and axial mesons receive a significant
contribution from confinement interactions of the order of $30\%$.


For some of the quark masses we have used, the resulting mass of the
$\rho$ and $a_1$  resonances turn out to be significantly smaller
than the threshold for decaying into three- and two- pions,
respectively. In the presence of confinement such hadrons would be
stable. However, in our IILM calculations we find a $\rho$ meson
width of $10$~MeV. Our interpretation of these results is that, in
the presence of chiral interaction only, the $\rho$ meson eventually
tends to dissociate into its constituents.

In the next session we present our phenomenological approach  to
extract information about hadron resonances from effective mass
plot. In session \ref{results} we  present and discuss the results
of our calculations. Session \ref{conclusions} is devoted to the
summary and conclusions.

\section{Resonances and the Effective Mass Plot}
\label{resonances}

\begin{figure}
\includegraphics[width=0.45\textwidth]{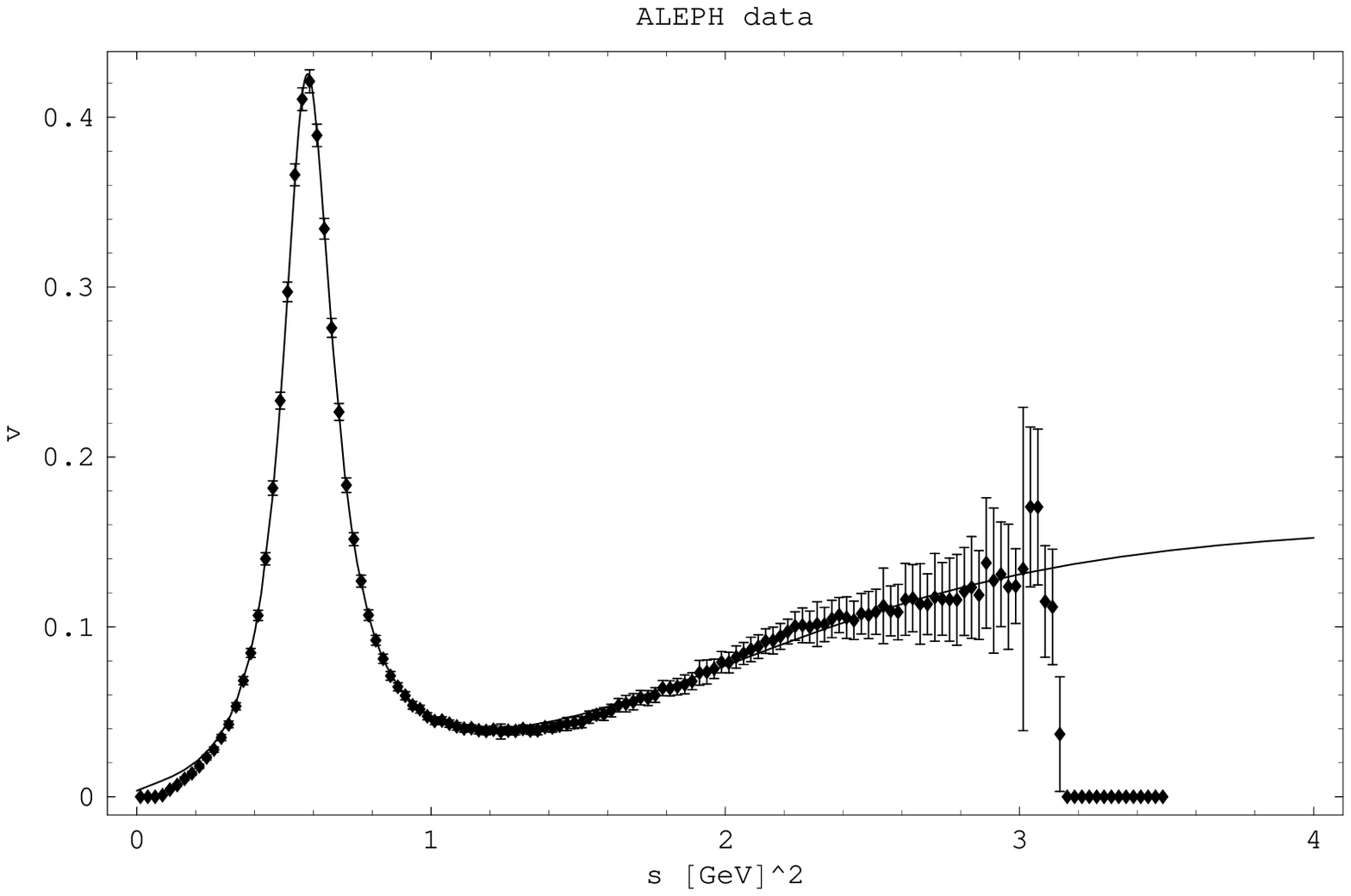}
\includegraphics[width=0.45\textwidth]{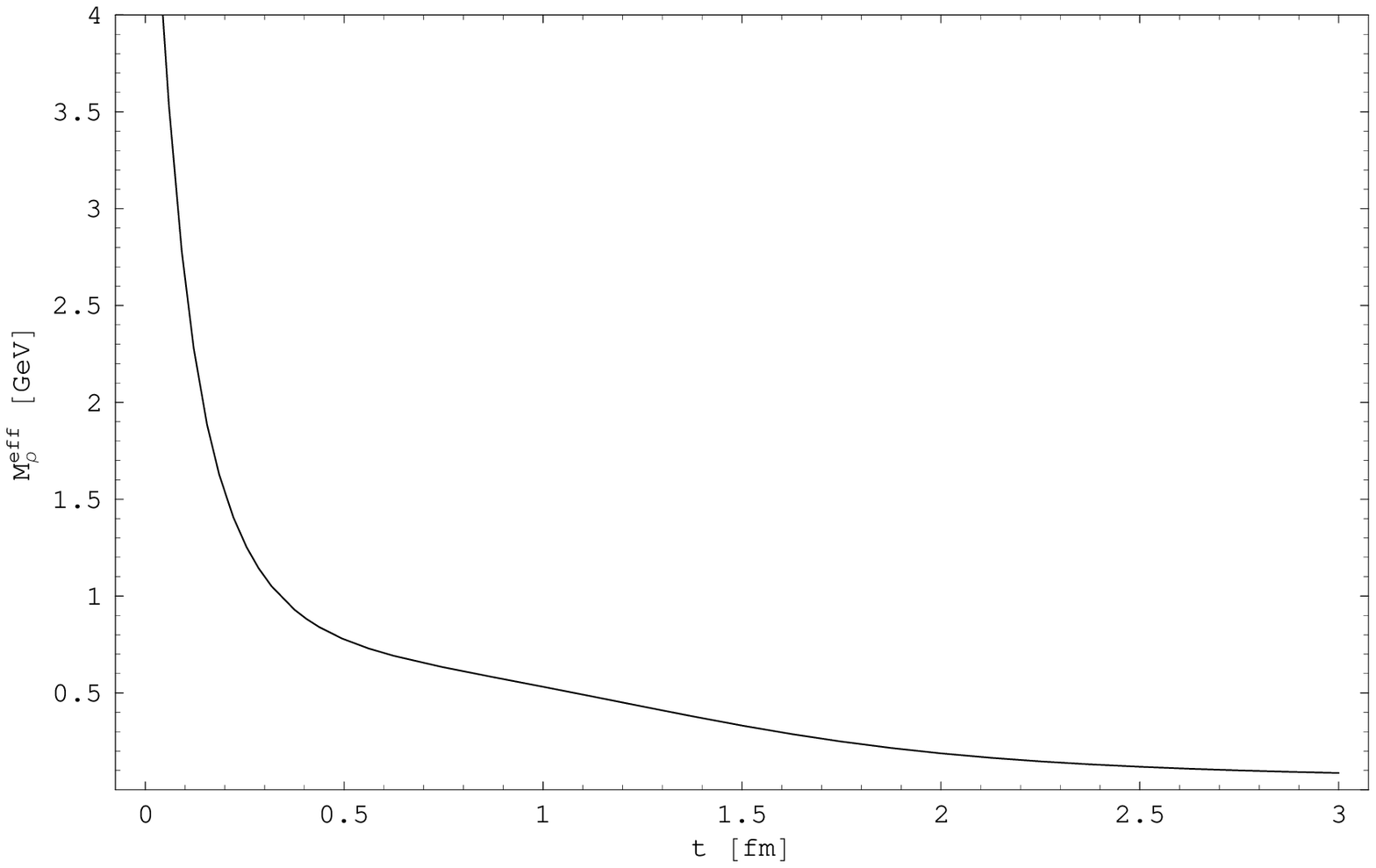}
\caption{Left panel: parametrization of the ALEPH Collaboration \cite{aleph:1998} data for the vector spectral density. Right panel: the corresponding effective mass plot}
\label{fig:aleph}
\end{figure}

In QCD the information about the hadron spectrum is encoded in the two-point correlation functions, defined as
\be
\Pi_H({\bf x}, \tau)= \langle 0| T [ J_H({\bf x}, \tau) J^\dagger_H({\bf 0}, 0)] |0\rangle.
\ee
$J_H({\bf x},\tau)$ is an overlap operator that creates states with the quantum numbers of the hadron $H$.
The lowest-dimensional overlap operators generating states with quantum numbers of $\pi$, $\rho$, $a_1$  mesons are:
\bea
J_{H}(x) &=& \bar{q}(x)~ \Gamma_H~ q(x),\\
\Gamma_\pi &=& \tau^+ i\gamma_5,~ \Gamma_\rho= \tau^+
\gamma_\mu,~\Gamma_{a_1}= \tau^+ i\gamma_5 \gamma_\mu. \eea In the
following we  consider the effective mass  $M^{eff}_H(\tau)$,
defined as \bea\label{eq:efmpr} M^{eff}_{H}(\tau) &=& \lim_{\Delta
\tau \to 0} \frac{1}{\Delta
\tau}~\ln\left[\frac{G_H(\tau)}{G_H(\tau+\Delta \tau)}\right], \eea
where $G_H(\tau)$ is the zero-momentum-projected hadronic two-point
function, \be G_H(\tau)= \int d^3 {\bf x}~\Pi_H({\bf x},
\tau)\label{GHdef}\,, \ee which can be written in the spectral
representation: \be
G_{H}(\tau)=\int\frac{\textrm{d}s}{2\sqrt{s}}\rho_{H}(s)e^{-\sqrt{s}\tau},
\label{GH} \ee where $\rho_{H}(s)$ is the spectral function.

In the large Euclidean time limit, the effective mass filters-out
the lowest singularity in the two-point function, i.e. the smallest
eigenvalue of the transfer matrix. If the lowest-lying state in a
given channel is a stable hadron, then the two-point function
develops a pole at the bound state mass below the threshold of the
branch-cut associated to multi-particle production. Then, the
effective mass asymptotically approaches the value of the mass of
the bound-state: \be \lim_{\tau\to \infty} M^{eff}_H(\tau) = M_H
\label{stable} \ee


In QCD the only bound states are pions and nucleons. In all the other channels, the two-pint functions displays branch-cut singularities only.

It is instructive to study the behavior of the effective mass plot if the spectral function $\rho_H(s)$ displays a  narrow resonance
with a finite-width, emerging above a continuum back-ground at small $s$ and converging to the asymptotic perturbative continuum, in the large $s$ limit. As a working example, we consider the effective mass plot for the vector meson channel. In this case, the spectral function can be extracted from the ALEPH Collaboration data \cite{aleph:1998} for $\tau$ decays in two pions. A reasonable
parametrization of such data can be constructed from a Breit-Wigner function for the $\rho$-meson resonance, supplemented by a term
simulating the  perturbative continuum~(see Fig.\ref{fig:aleph}, left panel)~\cite{shCF}: 
\be\label{eq:rhof}
    \rho_\rho(s)= C_1^{\rho}\frac{(\Gamma_{\rho}/2)^2}
    {(\Gamma_{\rho}/2)^2+(\sqrt{s}-m_{\rho})^2}+
    \frac{C_2^{\rho}}{1+\exp[(E_0-\sqrt{s})/0.2]}
\ee
The right panel shows the effective mass plot obtained from a
phenomenological parametrization of the two-point function, using
Eq.s (\ref{eq:efmpr}), (\ref{GH}) and (\ref{eq:rhof}). At small
Euclidean times, $\tau~\lesssim~0.4$~fm, the effective mass
$M^H_{eff}(\tau)$ drops exponentially. Such a rapid fall-off is due
to the exponential suppression of the perturbative continuum of
excitations induced by the propagation in the imaginary time.  At
larger Euclidean times, 0.5~fm~$\lesssim~\tau~\lesssim~2~$fm, the
effective mass displays a linear, nearly flat region. In this
regime, the spectral representation of $M^H_{eff}(\tau)$ is
dominated by the $\rho$-meson resonance peak. In fact, it is easy to
check that, in the limit of vanishing width, one recovers a
completely flat straight line, i.e. the familiar signature of a
stable bound-state. Eventually, at even larger $\tau$, the effective
mass slowly converges to the threshold energy for multi-particle
production\footnote{ Note that, since our simple phenomenological
parametrization (\ref{eq:rhof}) does not vanish below the two-pion
threshold, $s=\sqrt{2 m_\pi}$, the resulting effective mass
converges to $0$ in the asymptotically large Euclidean times.}. We
stress the fact that the effective mass plot analysis is much more
efficient than the corresponding point-to-point correlation function
study in distinguishing a resonance peak from a stable bound-state.

From this discussion it follows that it is in principle possible to
extract the width of the resonance from the effective mass plot in
the  intermediate Euclidean time region. We note that, in
(unquenched) lattice simulations, the stability of all  hadrons
except nucleon and pions depends on the size of the simulation box
and on the value of the pion mass used. For example, at large pion
masses $2 m_\pi > m_\rho$ the $\rho$-meson is a stable state,
because there is no phase-space available for decaying into two
pions. In this case, one can simply read-off its mass from the
plateau in the effective mass plot, at large Euclidean times.

On the other hand, for sufficiently small pion masses, the
phase-space for decaying into two-pions opens up and the
$\rho$-meson appears in the spectral density as a resonance. In this
case, the smallest eigenvalue of the transfer matrix filtered-out by
the propagation in imaginary time is related to the two-pion
$p$-wave state. Note however that, if the  periodic box is too
small, the quantization of momentum may shift the $p$-wave two-pion
state threshold above the $\rho$-meson mass.   As an example, let's
consider a  simulation performed in   a box with  size of $L=2.5~$fm
and with a pion mass of $500$~MeV. In this case,  the smallest
non-vanishing unit of momentum is $2\pi/L\simeq 500$~MeV, and the
threshold for decay into a two-pion $p$-wave state is at
$2\sqrt{m^2_\pi+ (\pi/L)^2}\simeq 1.1$~GeV which can be above the
$\rho$-meson mass.

Note that in our IILM calculations we do not have to worry about
effects related to  quantization of momentum, as we do not adopt
periodic boundary conditions. Instead, we choose simulation  boxes
which are sufficiently large for the integrand $\Pi_H({\bf x},\tau)$
in the momentum projection integral (\ref{GHdef}) to become very
small and negligible near the borders of the box\footnote{Note that
this is different from imposing Dirichlet boundary conditions, as we
do not impose  wave-functions or correlators to vanish at the border
of the box.}. Under such conditions, the lowest point in the
branch-cut for the $\rho$ meson two-point correlation function is
located at the threshold for two-pion production, i.e. $\sqrt{s}= 2
M_\pi$.

\begin{figure}
\includegraphics[width=0.32\textwidth]{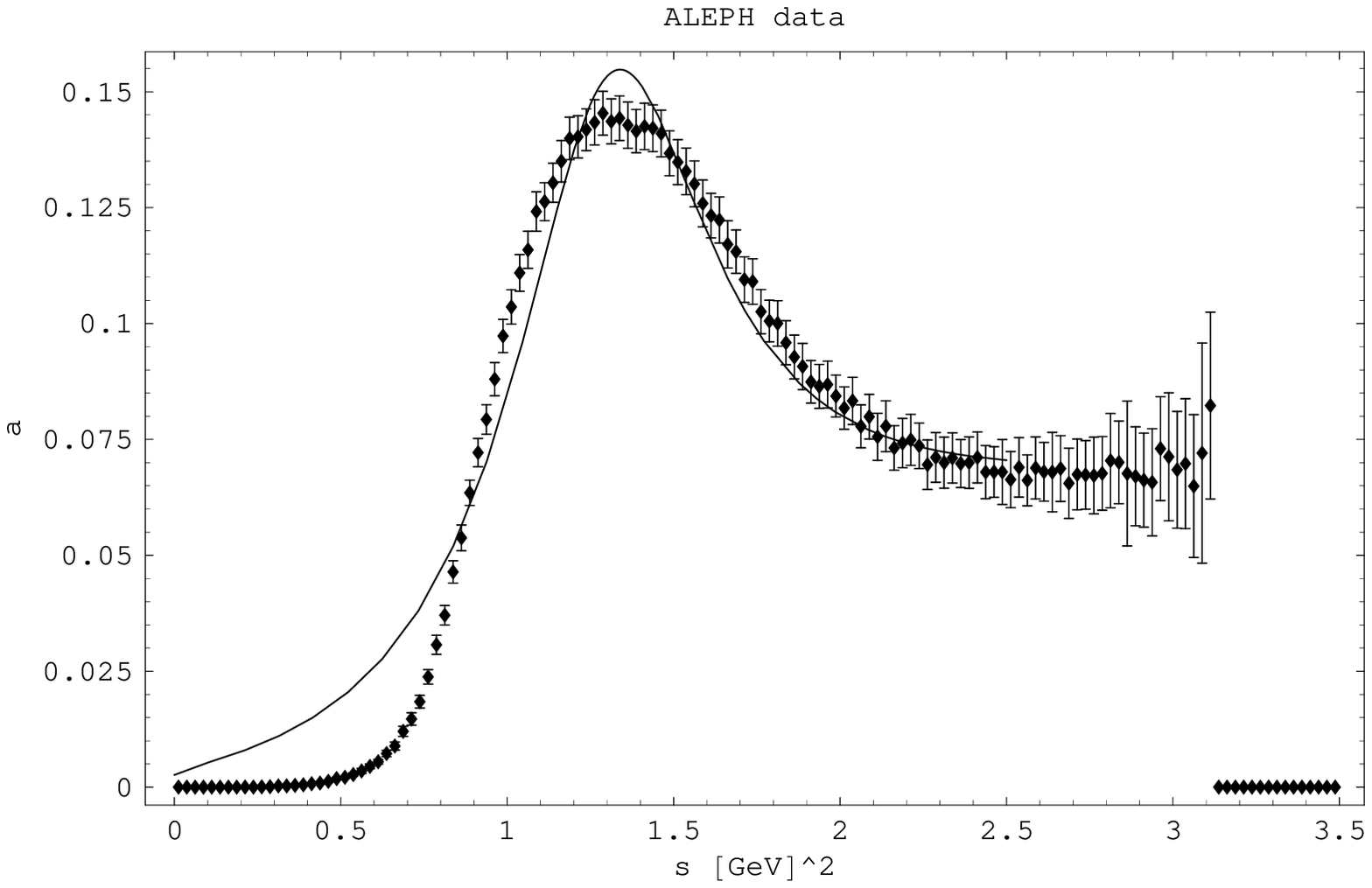}
\includegraphics[width=0.32\textwidth]{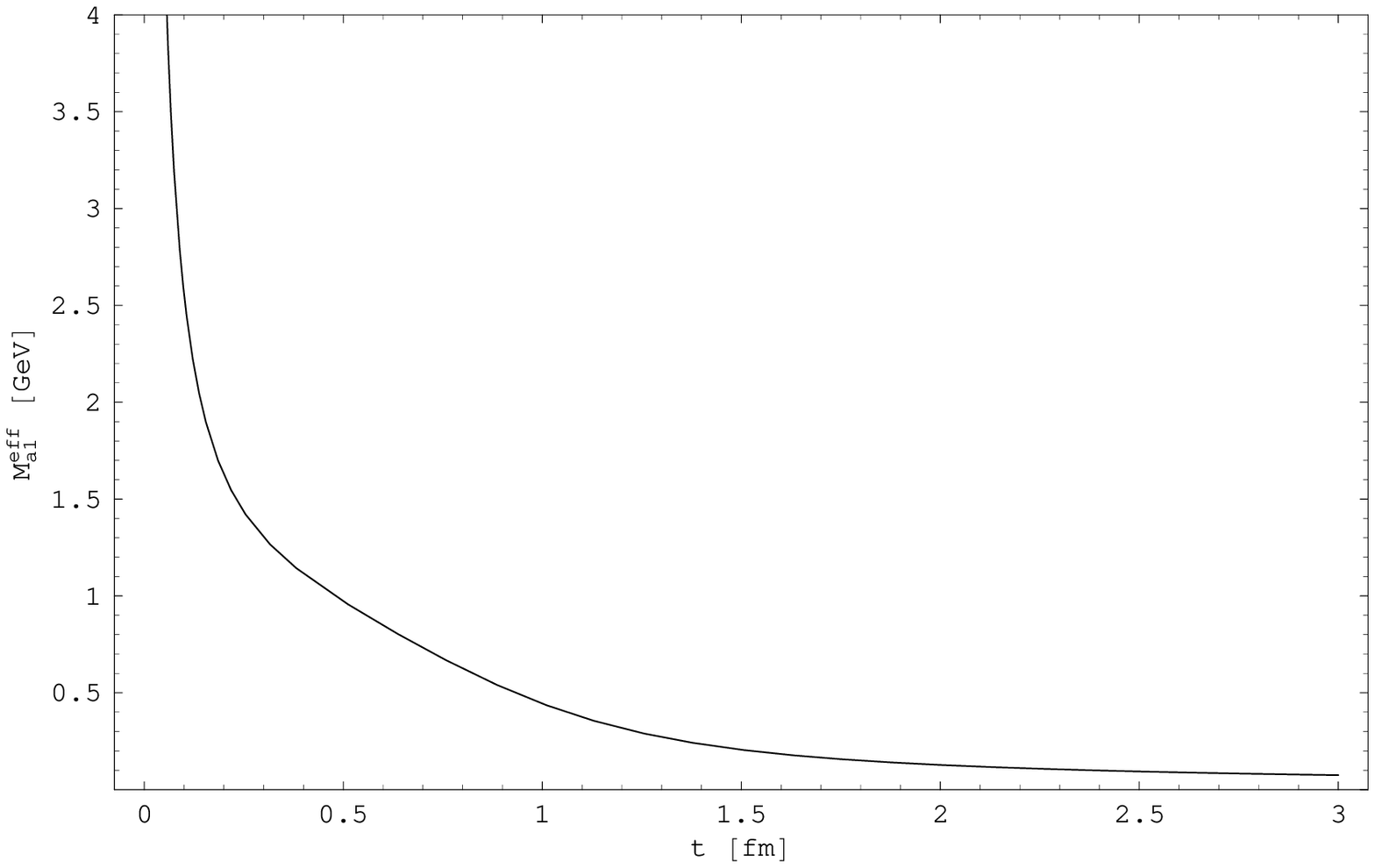}
\includegraphics[width=0.32\textwidth]{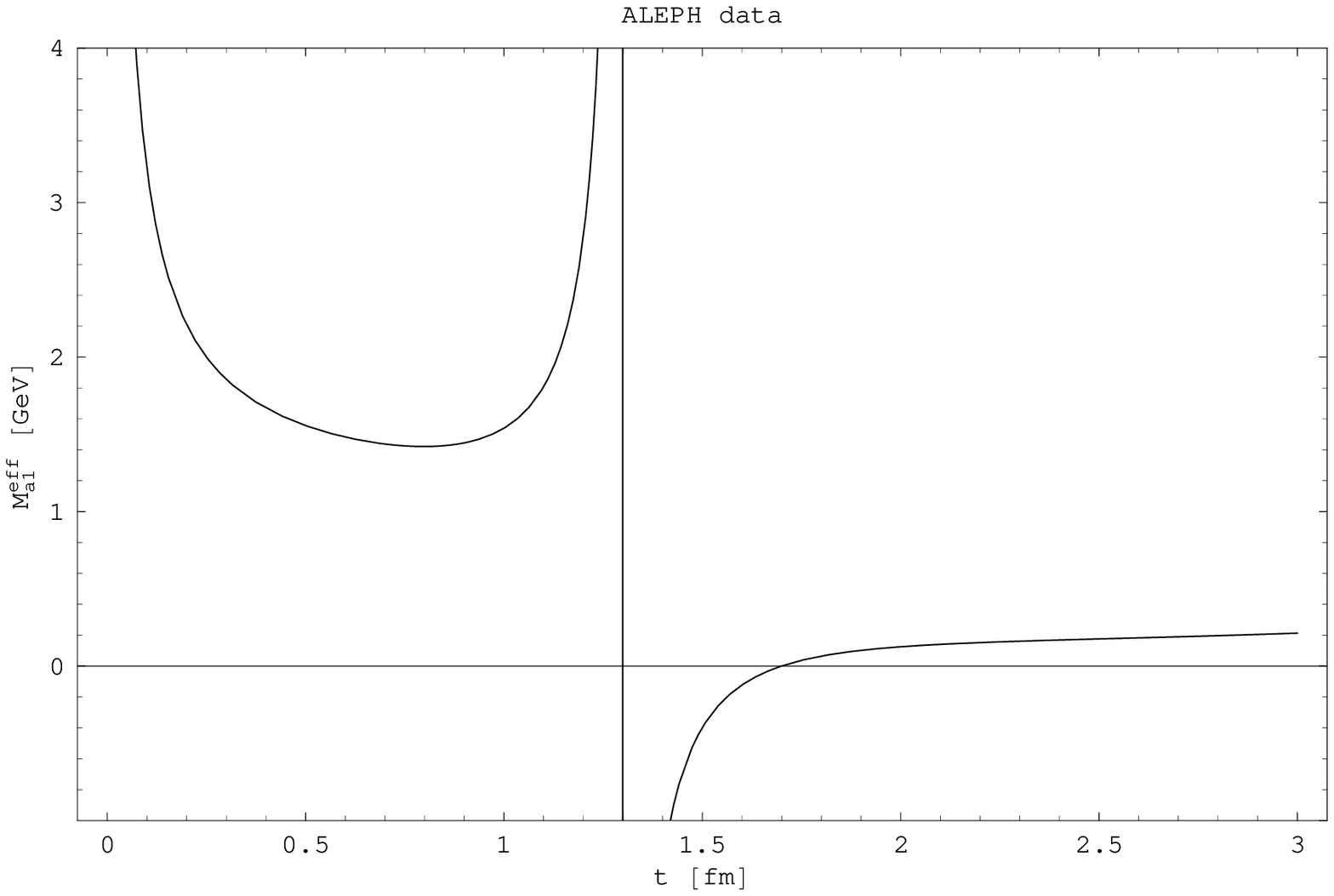}
\caption{Left panel: parametrization of the ALEPH Collaboration data \cite{aleph:1998} for $\tau$ decays in three pions using Eq.(\ref{eq:sfA}). Center panel:  the corresponding phenomenological effective-mass plot for the $a_1$ meson. Right panel: effective-mass plot for the $a_1$ meson with reduced width $\Gamma_{a_1}=0.03$ GeV}
\label{fig:empA1}
\end{figure}

Let us now discuss the axial-vector channel. In this case, the
hadronic current  has an overlap with both the pion state and the
$a_1$ resonance. A rough parametrization of the ALEPH Collaboration
data \cite{aleph:1998} for $\tau$ decays into three-pions  (see the
left panel of  Fig.~\ref{fig:empA1}) leads to the spectral
function~\cite{shCF}: \be\label{eq:sfA}
    \rho(s)= C_1^{a_1}\frac{(\Gamma_{a_1}/2)^2}{(\Gamma_{a_1}/2)^2+
    (\sqrt{s}-m_{a_1})^2}-f_{\pi}^2m_{\pi}^2\delta(s-m_{\pi}^2)+
    \frac{C_2^{a_1}}{1+\exp[(E_0-\sqrt{s})/0.2]},
\ee where the pion pole arises from the matrix element  $\langle 0|
J^\mu_5 (0)| \pi\rangle = i p_\mu f_\pi$. We note that the pion
contribution to this spectral function comes with an opposite sign
with respect to that of the $a_1$ resonance. Using (\ref{eq:sfA})
and dialing the physical value for $m_{\pi}$ and $f_{\pi}$, we
obtain the effective mass plot shown in the central panel of
Fig.\ref{fig:empA1}.
This plot displays a structure which is qualitatively similar to
that of the vector-meson channel. On the other hand, dramatic
differences emerge when the width of the $a_1$ is reduced (for
instance because of heavy pions in the spectrum). For example, if we
reduce $\Gamma_{a_1}$ by one order of magnitude we obtain the
effective mass plot shown in the right panel of Fig.\ref{fig:empA1}.
In the limit of stable $a_1$ (vanishing width), a singularity
develops  at the Euclidean time \be \tau=\frac{1}{M_{a_1}-M_\pi}
\log \left[\frac{\Lambda^2_{a_1}}{f_\pi^2 m_\pi^2}\right], \ee where
$\Lambda_{a_1}$ is the coupling  of the axial-vector current to the
$a_1$ state. This is a consequence of the  cancellation between the
contributions of the pion and axial-vector poles in the denominator
of Eq.~(\ref{eq:efmpr}).

In the next session we compare these phenomenological
representations  of the effective mass plot with the results of
calculations performed in the IILM.

\section{Results and Discussion}
\label{results}

\begin{table}[b!]
\caption{ $\pi$, $\rho$, $a_1$ masses (in GeV units) calculated in the IILM for different quark masses.}
\begin{center}\vspace{.2cm}
\begin{tabular}{llllllllllll}\hline\hline
    \rule{0pt}{3ex} $m_q$ & $$& $M_\pi$&$$ & $M_\rho$ &$$& $M_{a_1}$&$$&$\Gamma_\rho$&$$&$\Gamma_{a_1}$ \\ \hline
  \rule{0pt}{3ex}
    $0.02 $& $$& $ 0.30 \pm 0.04$& $$& $ 1.0\pm 0.1 $& $$& $  1.6 \pm 0.1$& $$& $\simeq 0.01 $& $$& $ < 0.02$\\
    $0.03 $& $$& $ 0.36 \pm 0.04$& $$& $ 0.9\pm 0.1 $& $$& $  1.6 \pm 0.1$& $$ &  $\lesssim 0.01 $& $$& $ < 0.03$\\
    $0.05 $& $$& $ 0.46 \pm 0.04$& $$& $ 1.0\pm 0.1 $& $$& $  1.7 \pm 0.1$& $$ & $\simeq 0.05 $& $$& $ < 0.01$\\
    $0.07 $& $$& $ 0.53 \pm 0.04$& $$& $ 1.0\pm 0.1 $& $$& $  1.7 \pm 0.1$& $$ & $\simeq 0.05 $& $$& $ < 0.01$\\
    $0.09 $& $$& $ 0.60 \pm 0.04$& $$& $ 0.9\pm 0.1 $& $$& $  1.8 \pm 0.2$& $$ & $\simeq 0.05 $& $$& $ < 0.01$\\
\hline\hline
\end{tabular}
\end{center}
\label{table:mes}
\end{table}

In the IILM,  hadronic correlation functions are evaluated by means
of Monte Carlo averages over instanton ensamble configurations. The
only phenomenological parameters of the model are the instanton
average size $\bar{\rho}=0.33~$fm and the dimensionless strength of
the instanton-antiinstanton bosonic short-distance repulsion (for a
concise review of this model see \cite{IILMchpt}, for an extended
treatment see \cite{rev}). In the present calculations we used
five sets of ensemble configurations, corresponding to quark masses
ranging from 20 to 90~MeV and we estimated statistical errors using
jackknife technique, with bin size of 10 configurations. In order to
isolate the instanton-induced chiral interactions, we have adopted
the so-called zero-mode approximation, in which the part of the
quark propagator  which does not receive contribution from the
instanton zero-modes has been replaced by a free propagator (for
further details, see the discussions in \cite{rev} and
\cite{IILMchpt}). For comparison, we have performed the same
calculations also including the non-zero-mode part of the propagator
and we have not found significant differences, hence
instanton-induced correlations not associated to the chiral
zero-mode zone play only a marginal role.

\begin{figure}
        \centering
        \subfigure{\includegraphics[width=.38\textwidth]{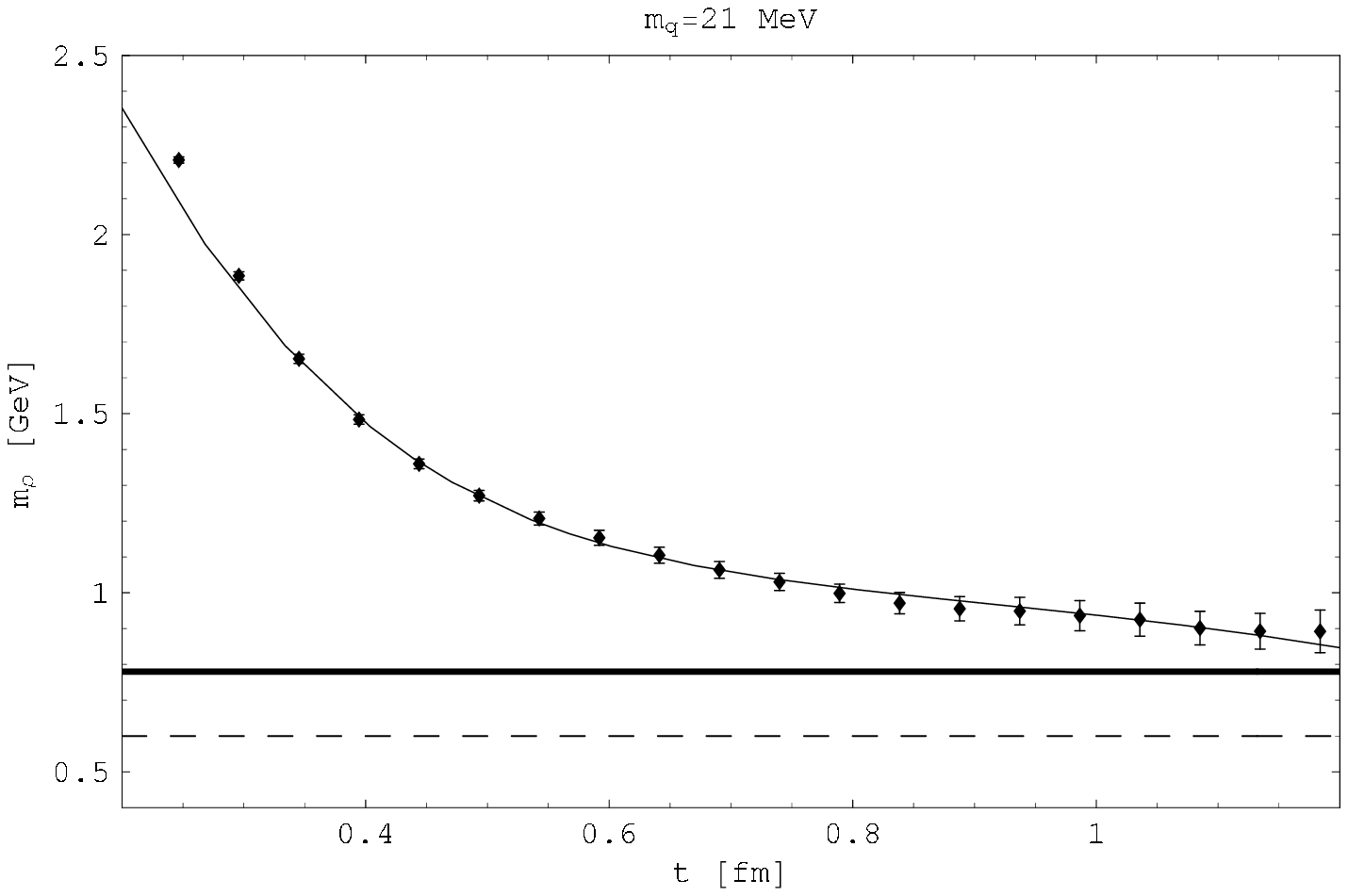}}\hspace{5mm}%
        \subfigure{\includegraphics[width=.38\textwidth]{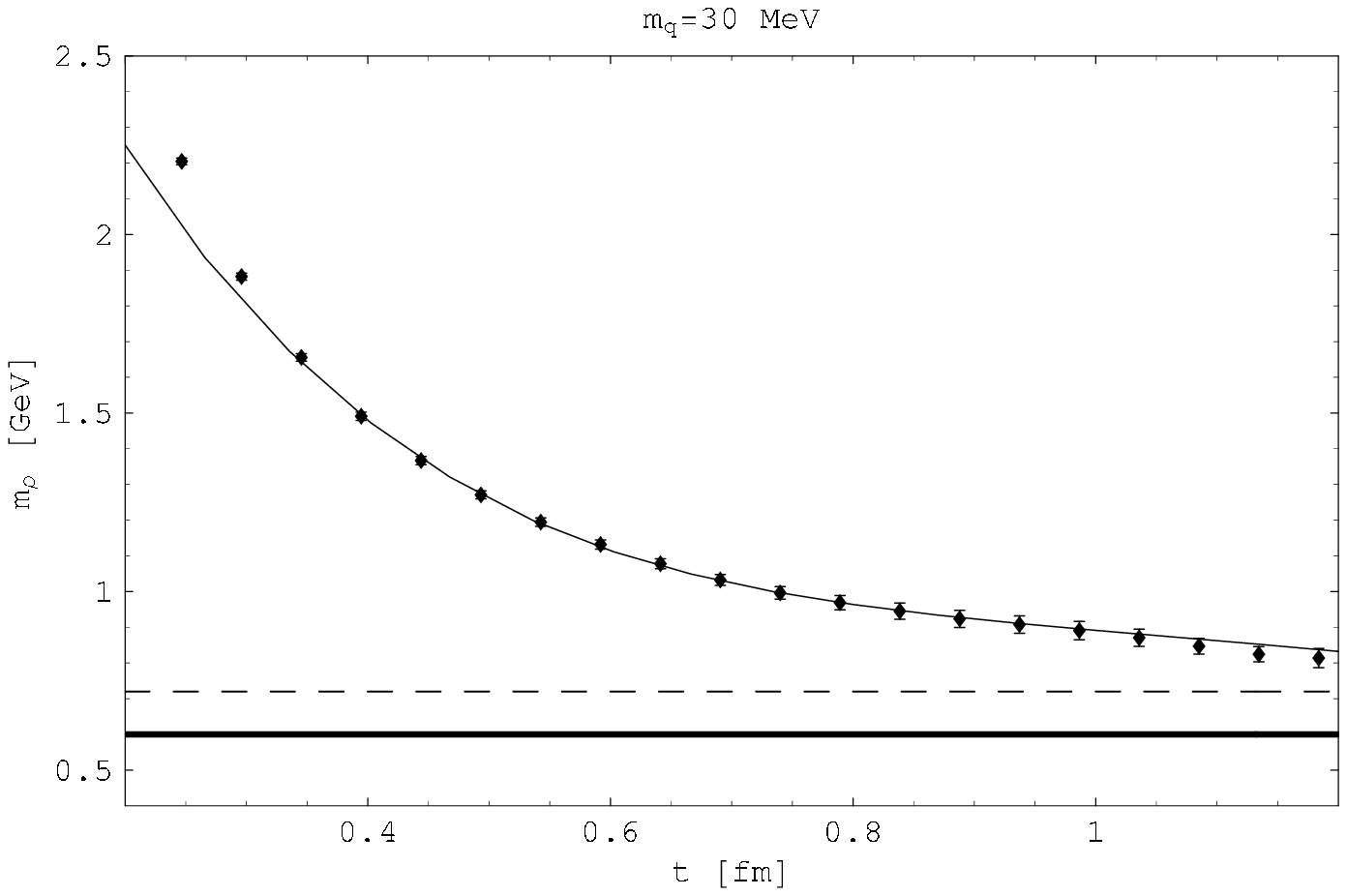}}
        \subfigure{\includegraphics[width=0.38\textwidth]{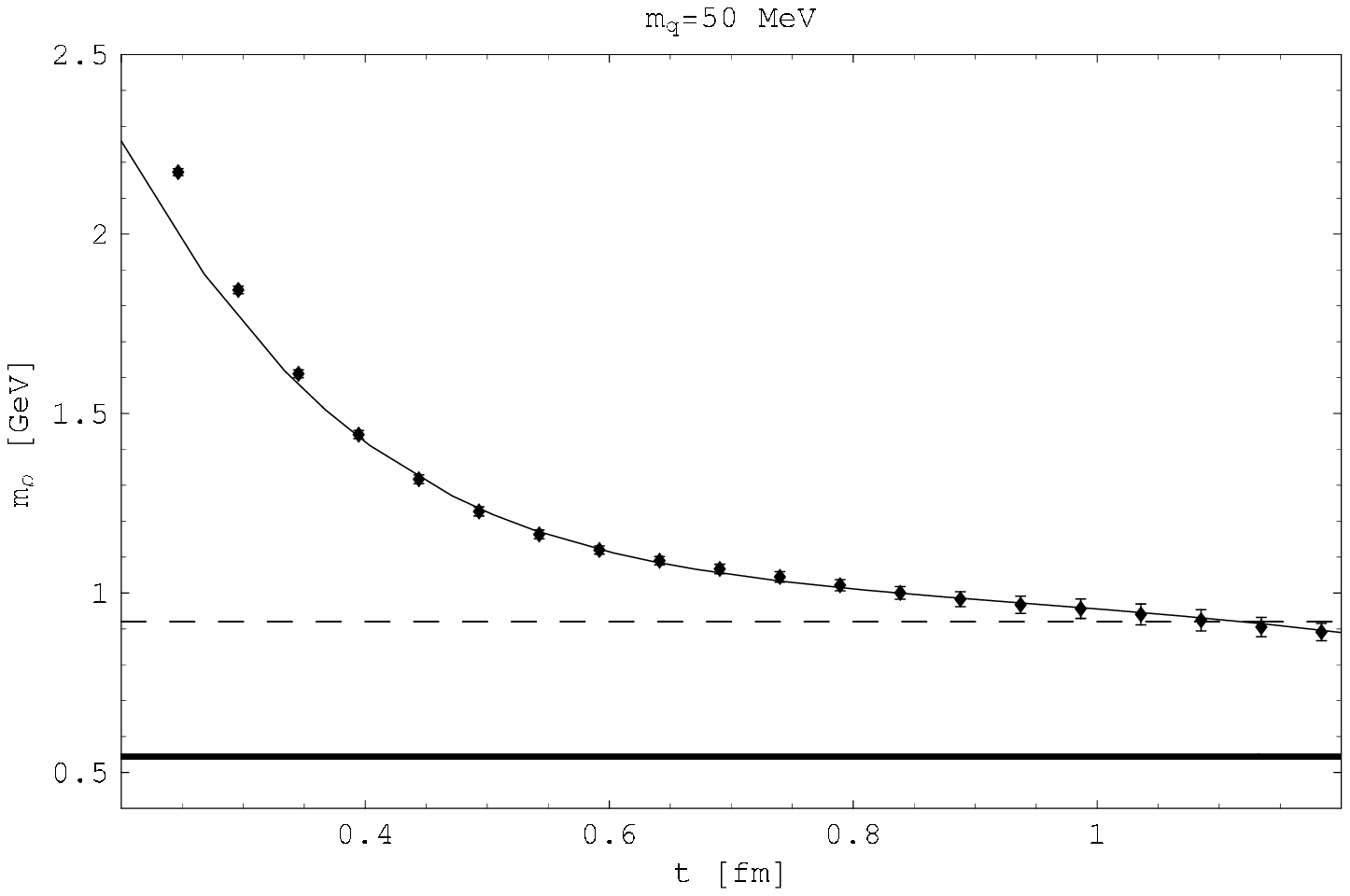}}\hspace{5mm}%
        \subfigure{\includegraphics[width=0.38\textwidth]{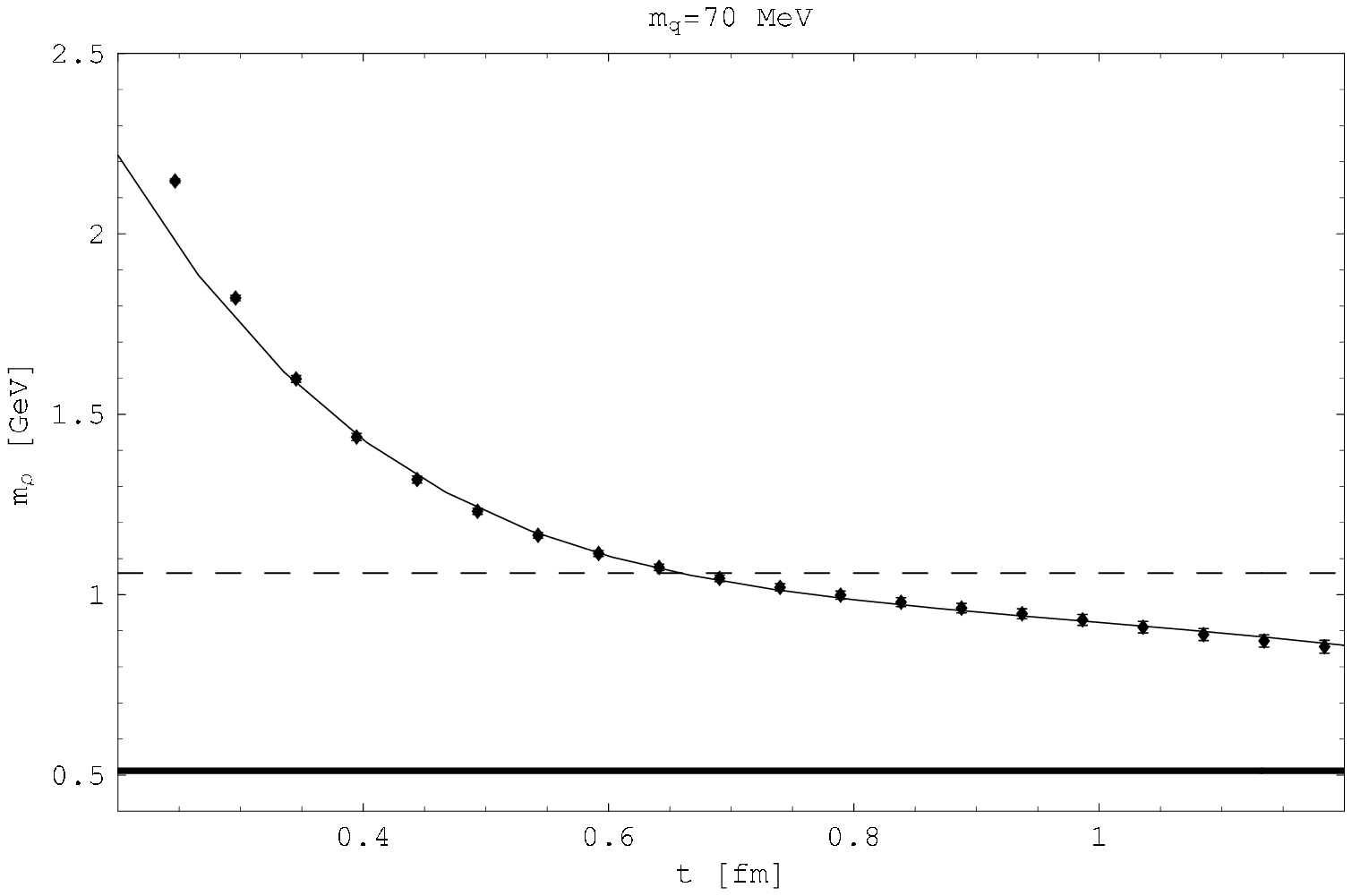}}
        \subfigure{\includegraphics[width=0.38\textwidth]{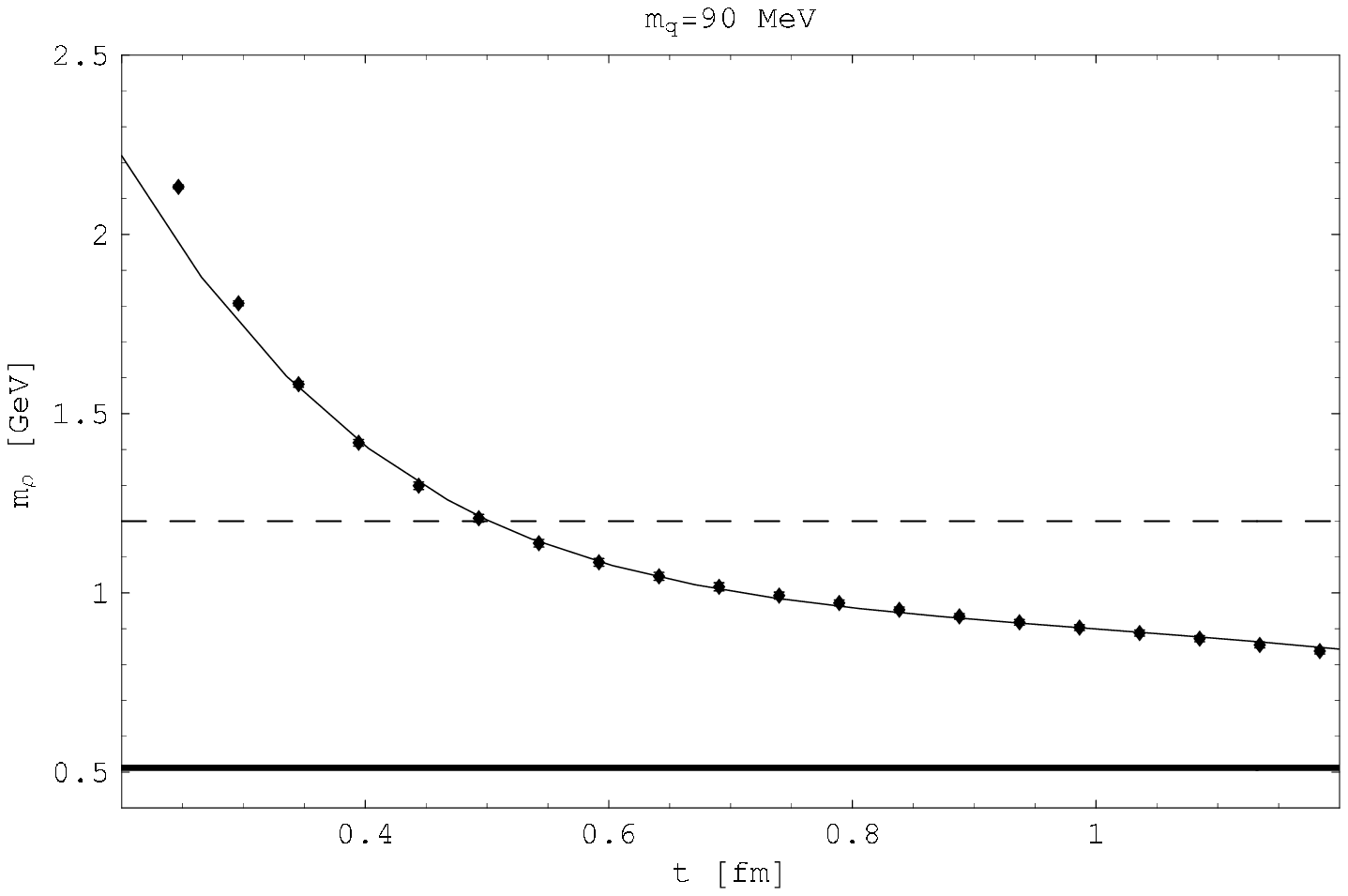}}\hspace{5mm}%
      \caption{ Effective mass plots in the $\rho$ meson channel evaluated in the IILM at different quark masses and compared to the phenomenological parametrization (solid line). The dashed straight  line represents the expected asymptotic plateau if the $\rho$ meson decays into two pion, while the solid straight line represents the expected asymptotic plateau if the $\rho$ meson decays into two constituent quarks with masses estimated from the ILM calculation in \cite{musakh}.}\label{fig:rho}
\end{figure}

Our results for the calculation of the $\rho$ meson effective mass
for several quark masses are shown in Fig.\ref{fig:rho}, where they
are compared with the best fit obtained from the phenomenological
representation of the spectral function (\ref{eq:rhof}). The IILM
points can be very well interpolated throughout the entire region
$0.4~$fm~$\lesssim\tau\lesssim~1.2$~fm. On the other hand, it should
be noted that a fit of these points using a spectral function which
does not account for a narrow meson resonance would be inconsistent
with our IILM points. The $\rho$ meson contribution is needed to
explain the nearly flat behavior of the effective mass for
$\tau>~0.6$~fm.

\begin{figure}
        \centering
        \subfigure{\includegraphics[width=.38\textwidth]{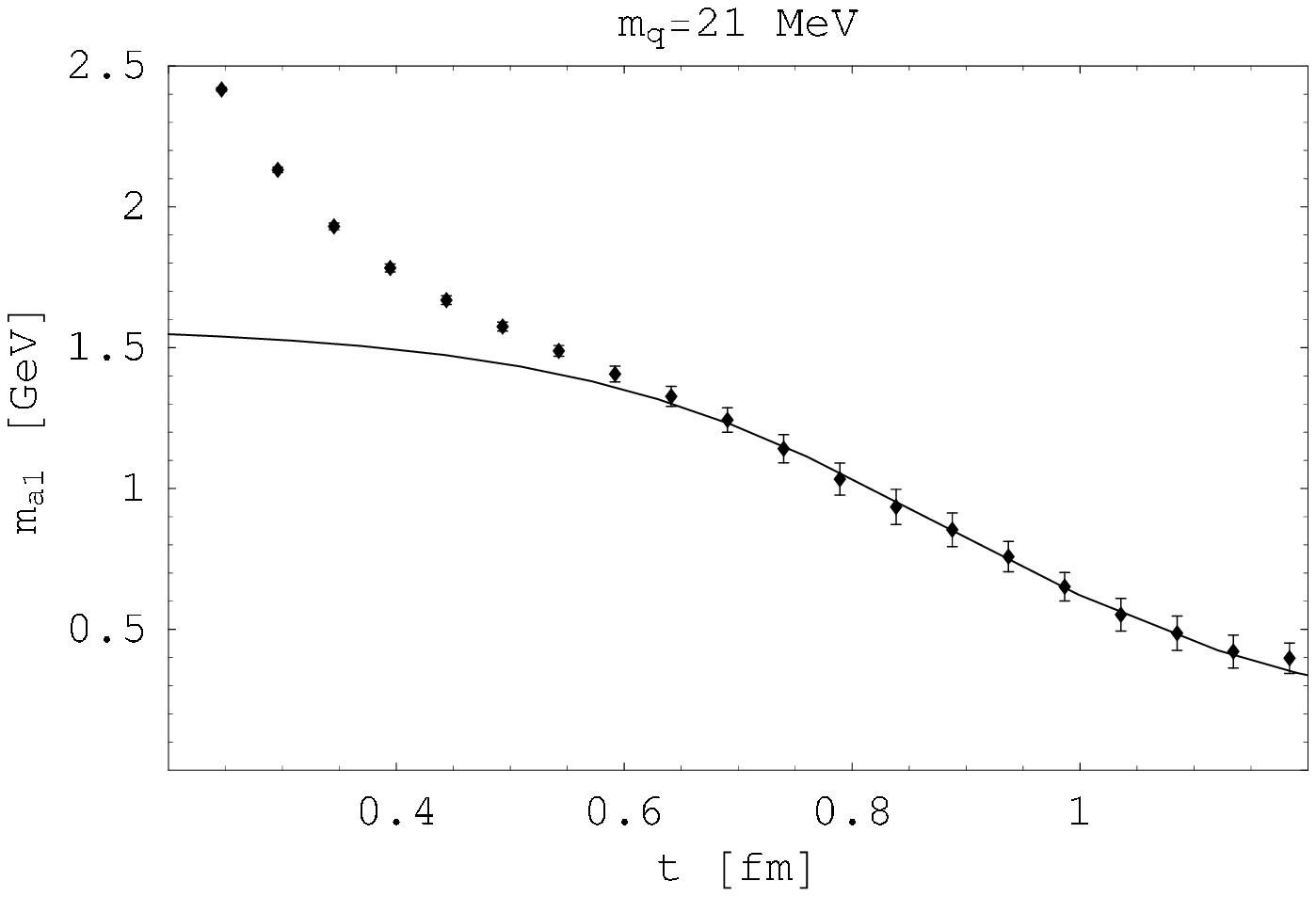}}\hspace{5mm}%
        \subfigure{\includegraphics[width=.38\textwidth]{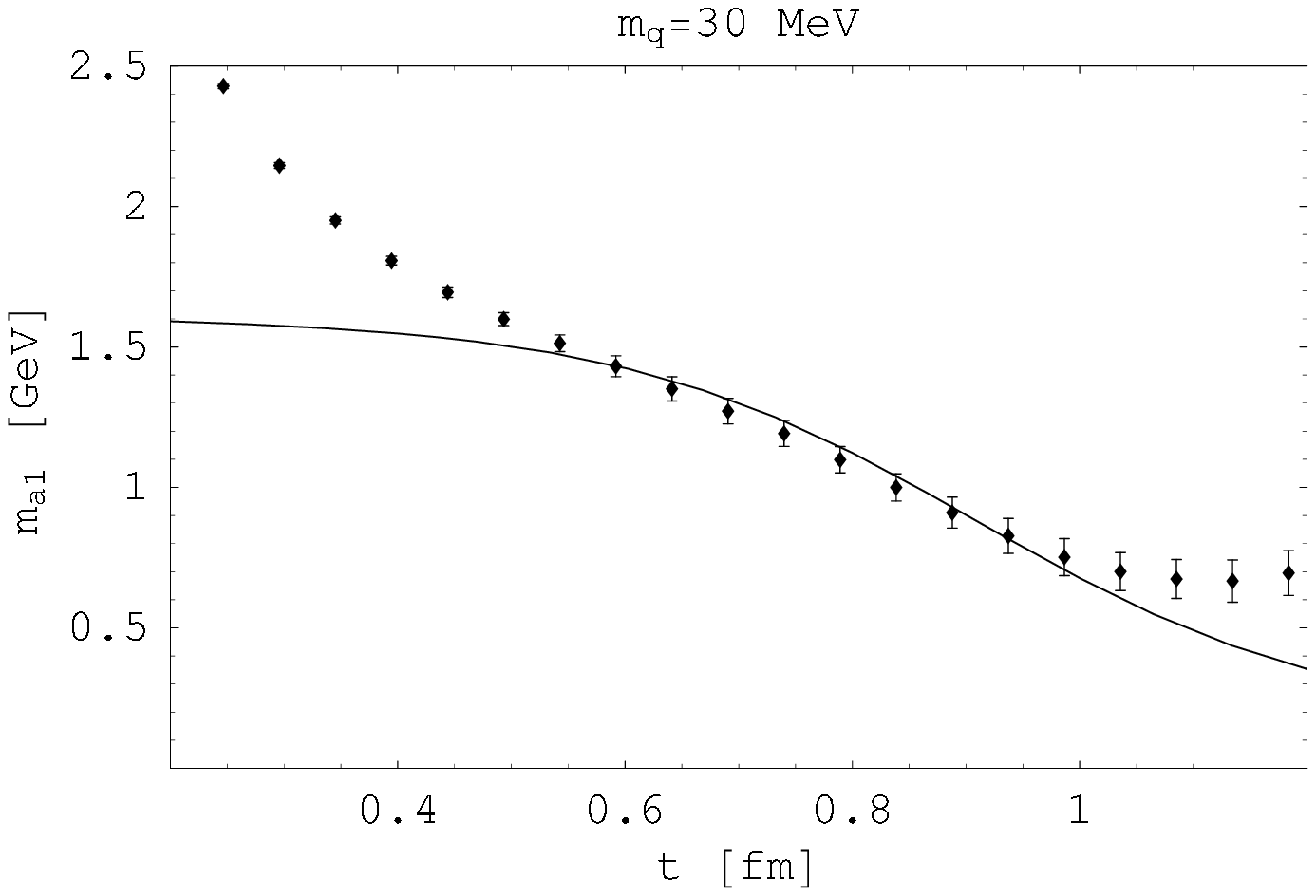}}
        \subfigure{\includegraphics[width=0.38\textwidth]{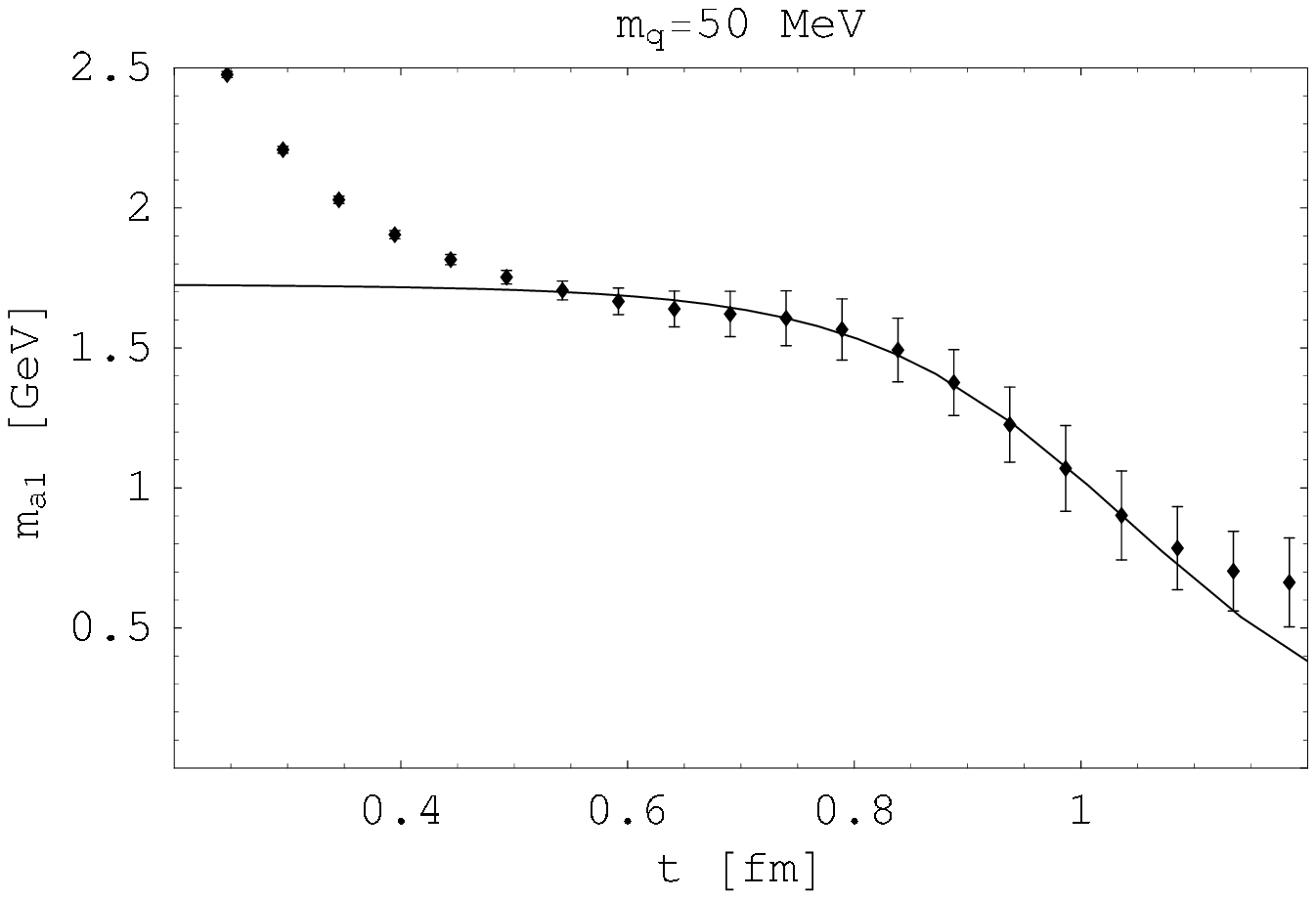}}\hspace{5mm}%
        \subfigure{\includegraphics[width=0.38\textwidth]{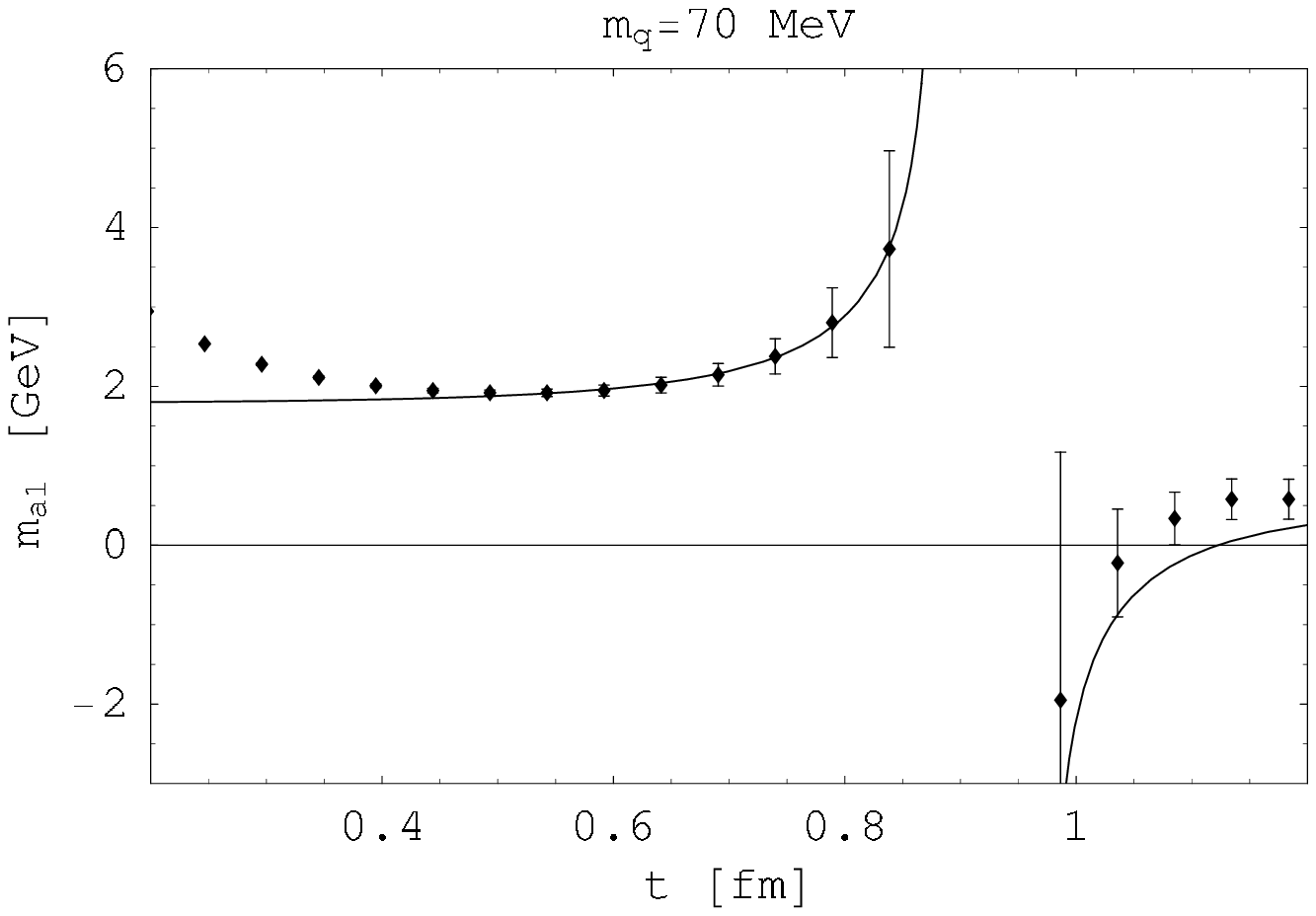}}
        \subfigure{\includegraphics[width=0.38\textwidth]{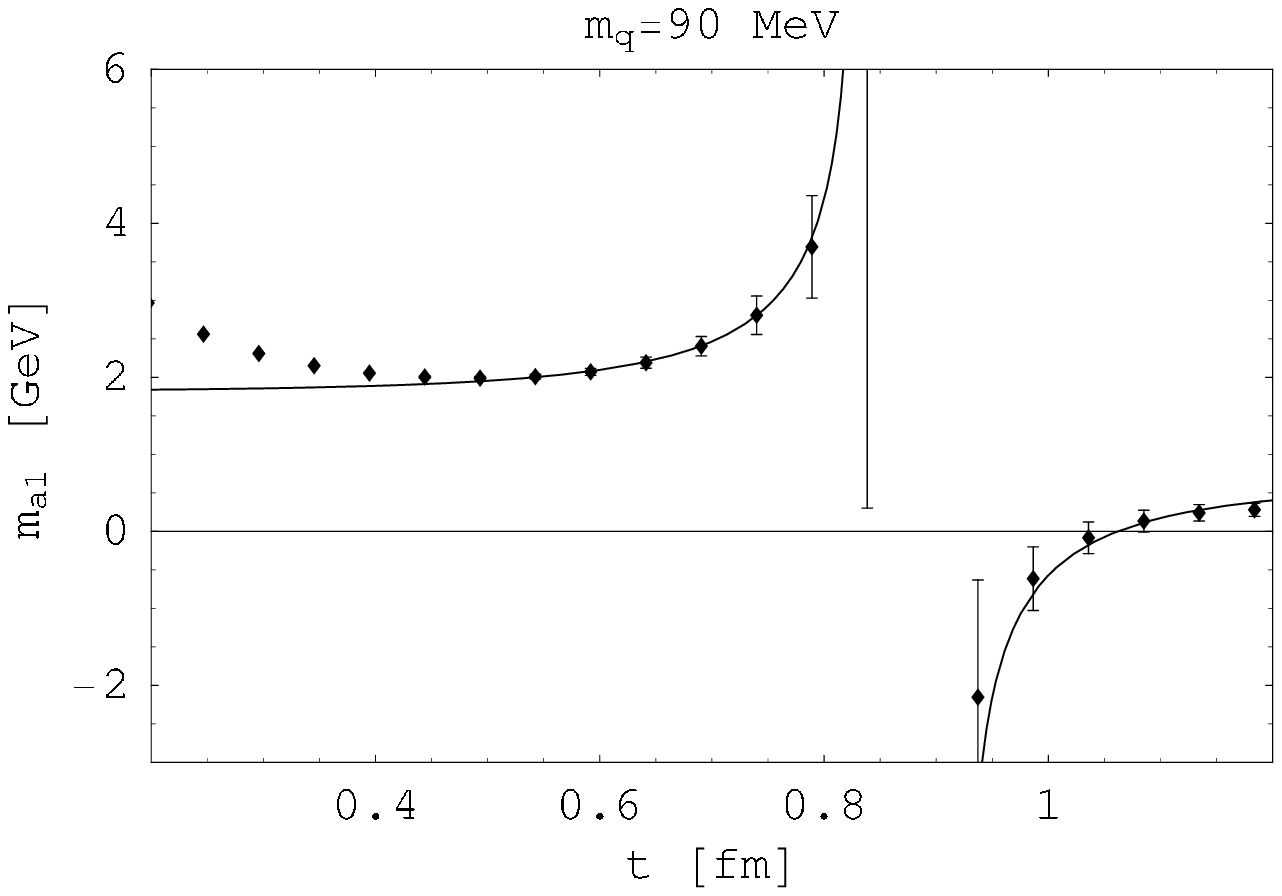}}\hspace{5mm}%
      \caption{ Effective mass plots in the $a_1$ meson channel, evaluated  in the IILM at different quark masses and compared to the phenomenological parametrization (solid line). The latter does not include the contribution from the perturbative continuum (see discussion in the text).}\label{fig:a1}
\end{figure}

Our results for the calculation of the $a_1$ meson effective mass
for several quark masses are shown in Fig.\ref{fig:a1}, where they
are compared with the best fit obtained from the phenomenological
representation of the spectral function (\ref{eq:sfA}). In order to
reduce the number of free fitting parameters, we have chosen to
neglect the contribution of the continuum, and to restrict the fit
of the IILM points to the region $\tau>0.6$~fm. In addition, we have
used the value for the pion mass and decay constant calculated in
the IILM in \cite{IILMchpt}. The qualitative behavior predicted in
the previous session is very well reproduced by our model. In
particular, we observe that the singularity arising from the
cancellation  of the $\pi$ and $a_1$ contribution to the two-point
correlator is clearly developed. This result provides a clean
evidence that both the pion and  the $a_1$ meson exist in the
instanton vacuum. We note that, in this channel, the simple
parametrization (\ref{eq:sfA}) of the ALEPH Collaboration data
\cite{aleph:1998} is quite poor in the low $s$ region. This is
presumably the source of the small discrepancy observed for some
quark masses, at the largest Euclidean times.

\begin{figure}
        \centering
        \subfigure{\includegraphics[width=.38\textwidth]{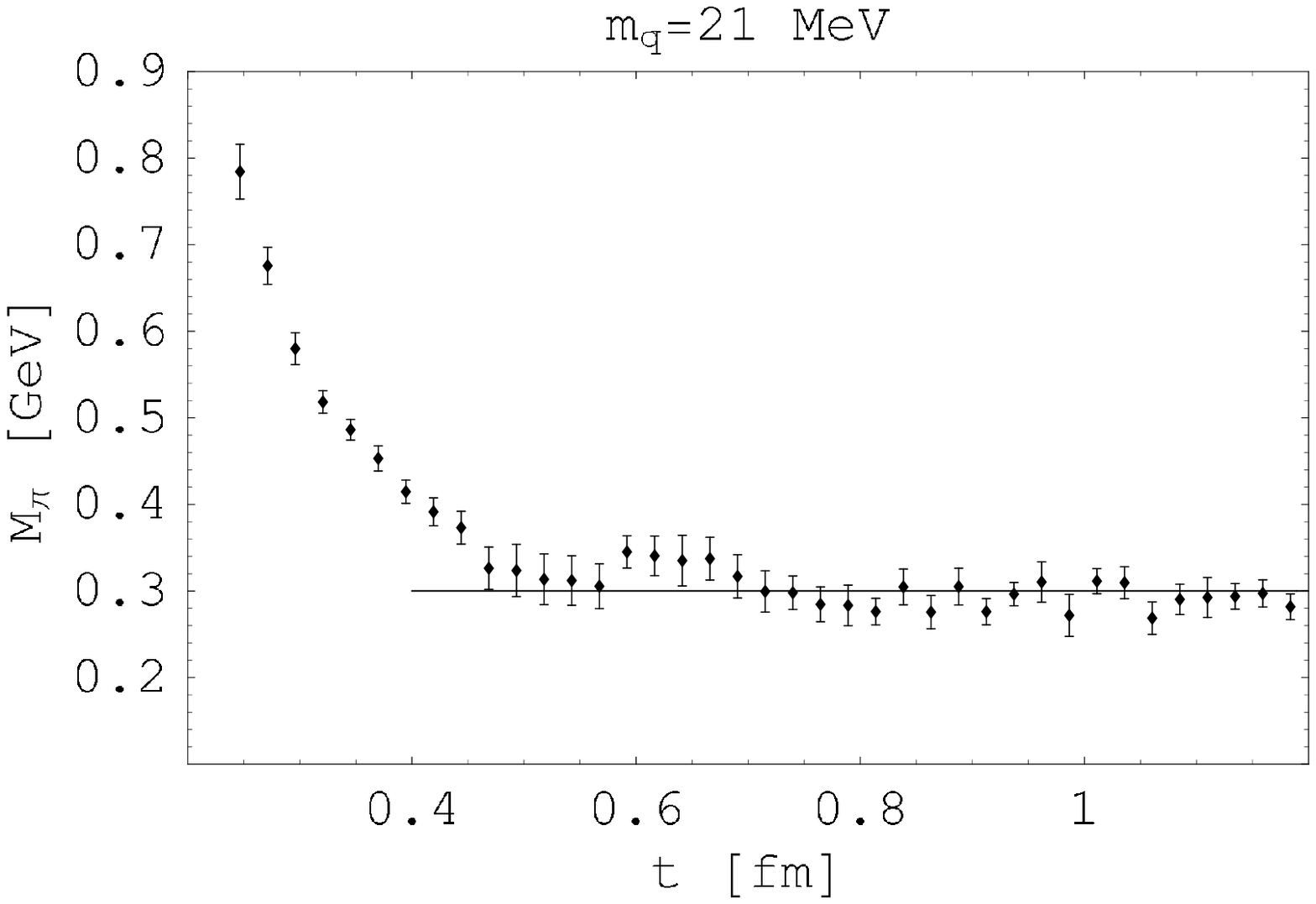}}\hspace{5mm}%
        \subfigure{\includegraphics[width=.38\textwidth]{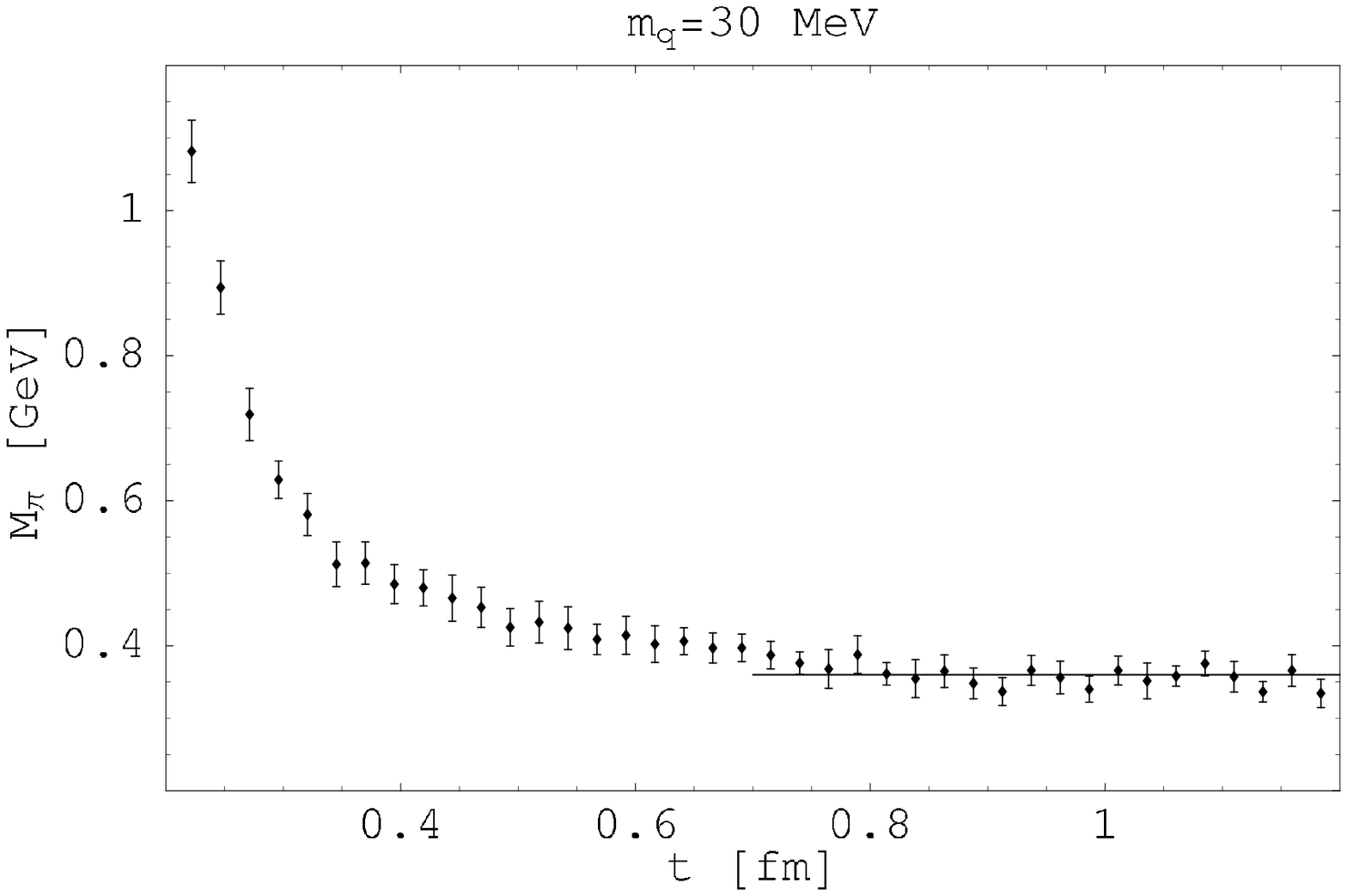}}
        \subfigure{\includegraphics[width=0.38\textwidth]{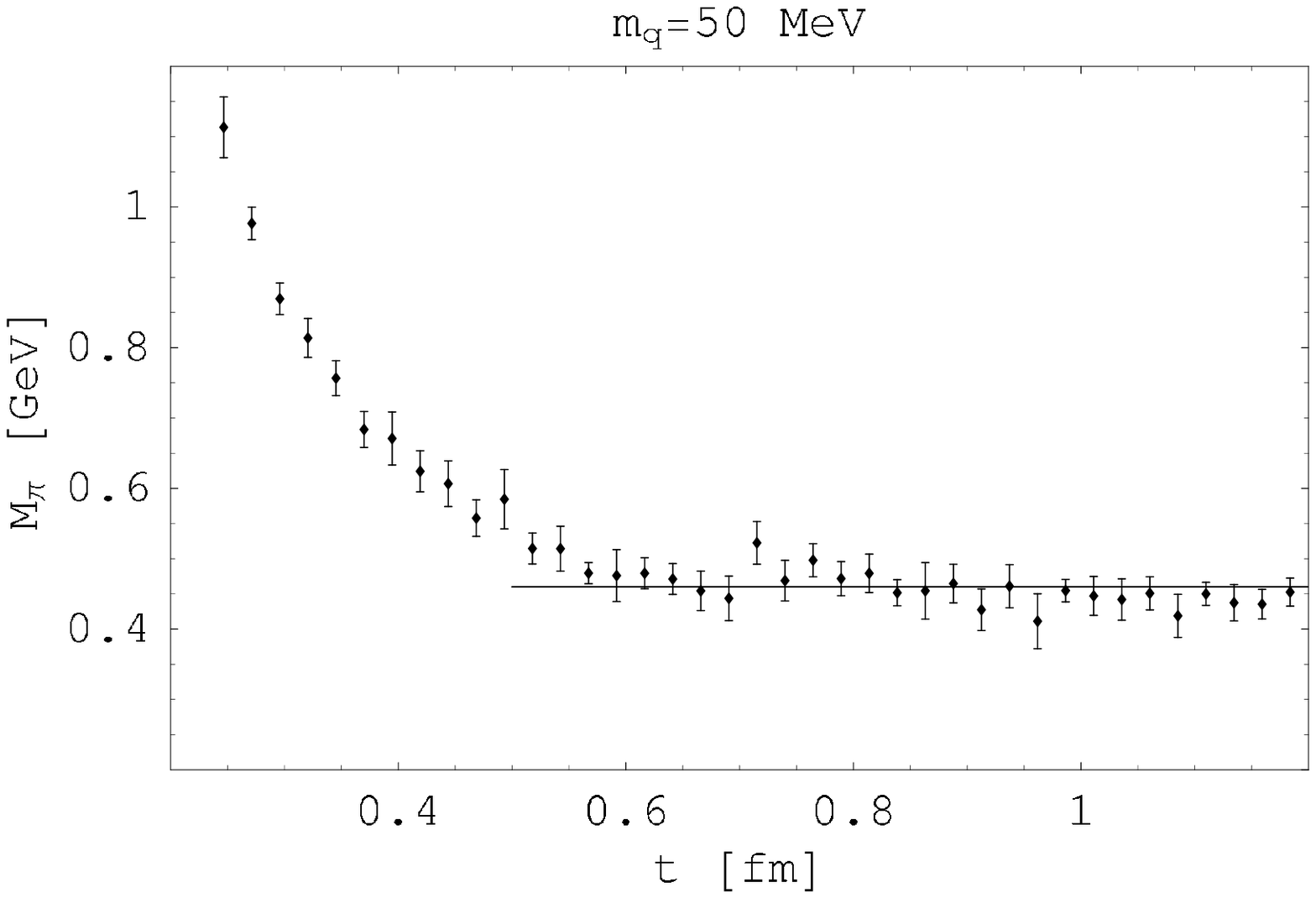}}\hspace{5mm}%
        \subfigure{\includegraphics[width=0.38\textwidth]{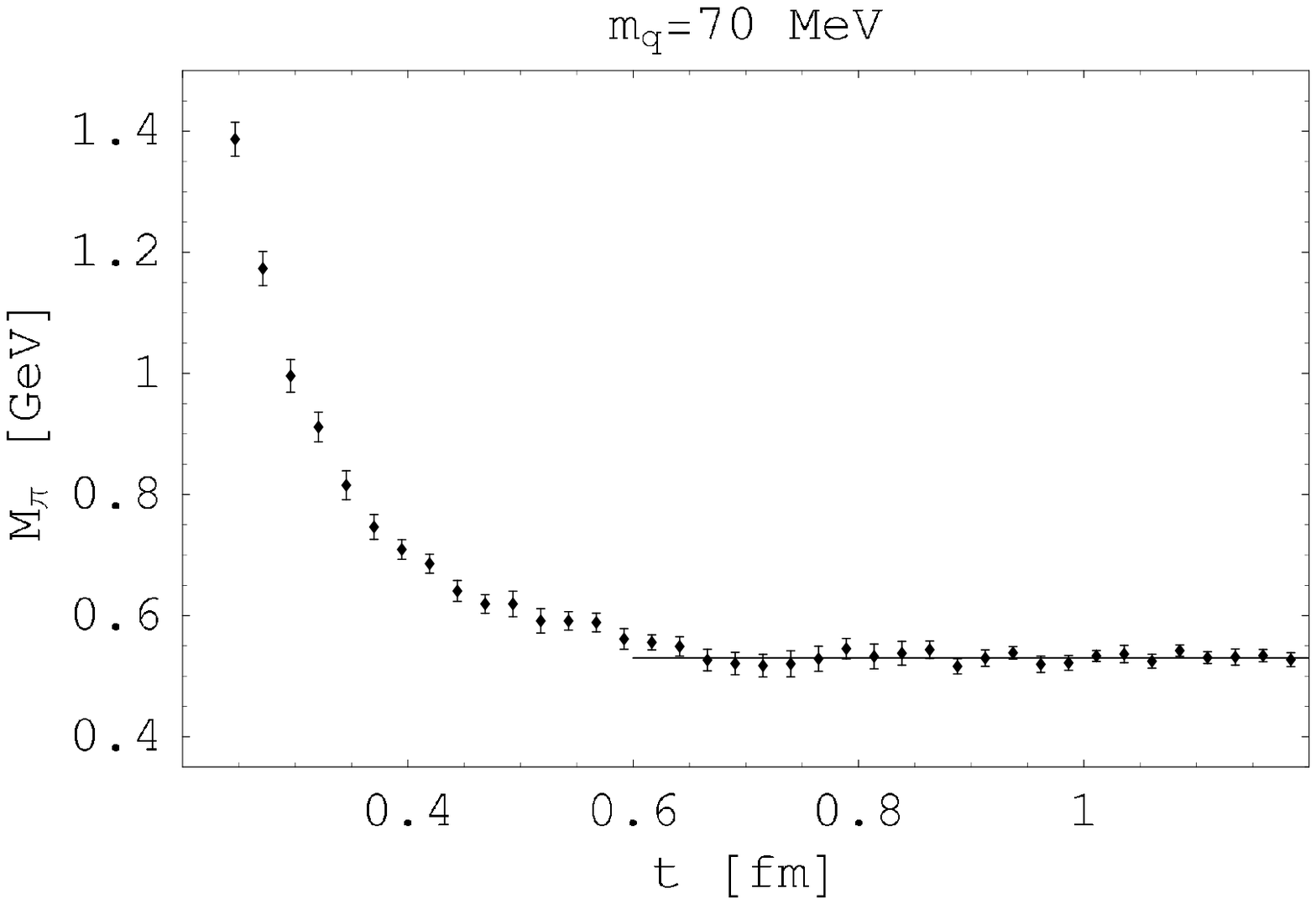}}
        \subfigure{\includegraphics[width=0.38\textwidth]{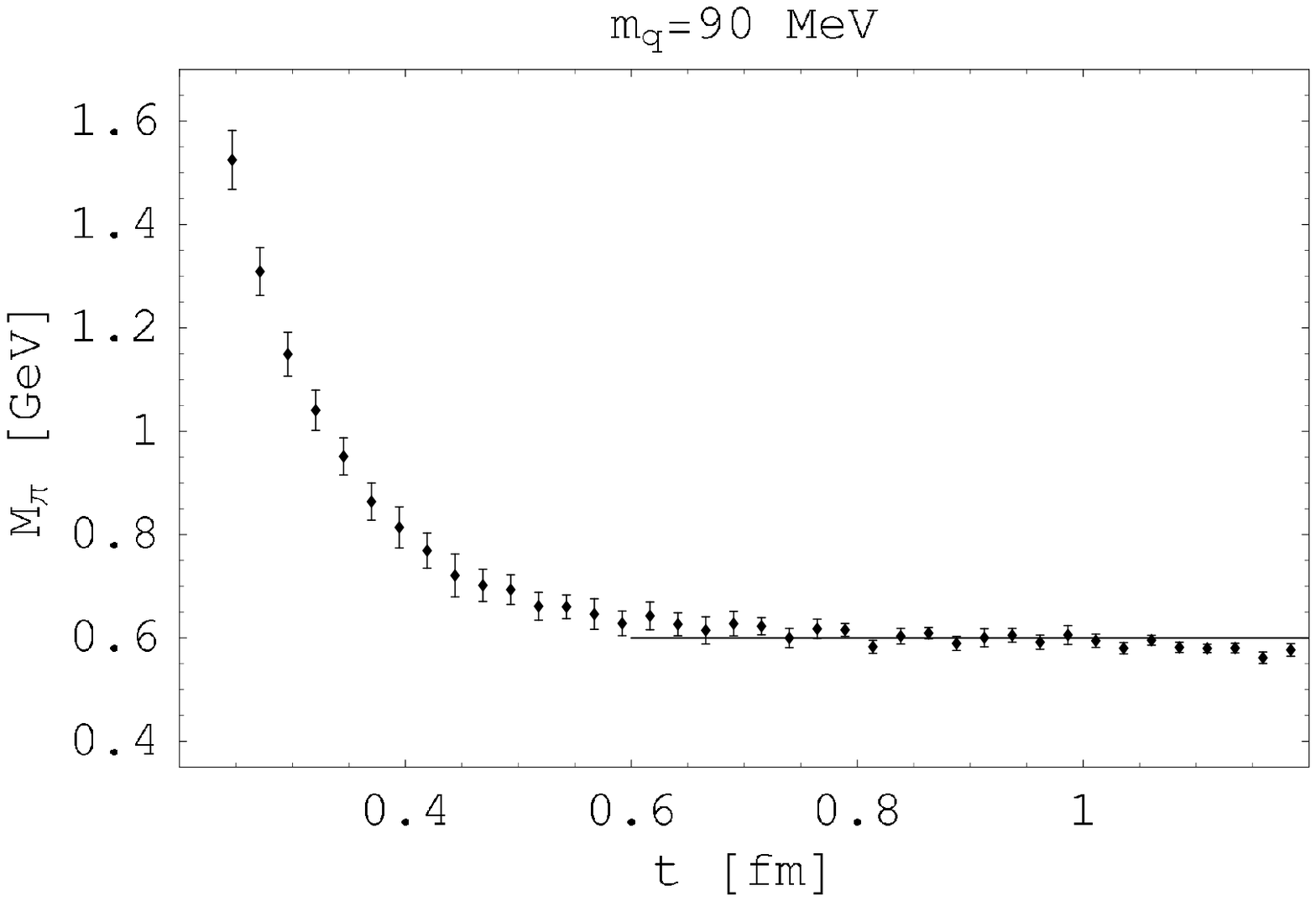}}\hspace{5mm}%
\caption{Effective-mass plots for the pion evaluated in the  IILM in \cite{IILMchpt} }
\label{fig:pionemp}
\end{figure}

For comparison, in Fig.\ref{fig:pionemp} we show our effective mass
plots for the pion, which were obtained in \cite{IILMchpt} and
display a completely flat behavior at large Euclidean times.


The complete list of $\rho$ meson and $a_1$ masses extracted from
the fit of  the effective mass plot are summarized in
Table~\ref{table:mes}. In general the masses obtained in the IILM
are about $30\%$ larger than  the  corresponding experimental
values. This discrepancy suggests that about $1/4$ of the resonance
mass is due to correlations which are not related to chiral symmetry
breaking.  This hypothesis is confirmed by the fact that the chiral
asymmetry, \be \chi= \frac{m_{a_1}-m_{\rho}}{m_{a_1}+m_{\rho}} \ee
which has been suggested as a parameter quantifying the contribution
of chiral forces to hadron mass splittings~\cite{chiasymm}, is
remarkably well reproduced in this model. At the lightest quark mass
we find $\chi^{IILM}\simeq 0.2$, which should be confronted with the
experimental value $\chi^{exp.}=0.23$.

Let us now look in more detail on the mechanism of meson decay and
we focus on the $\rho$ meson in the IILM. The effective mass plots
in the vector meson channel calculated at different quark masses are
consistent with a small width spanning from $10$~MeV to $50$~MeV.
Interestingly,  the vector meson is always unstable in this model,
even at quark masses for which  there is no phase-space for decaying
into pions, since $2 M_\pi> M_\rho$. A natural explanation of this
fact is that, in the absence of confinement, these hadrons decay
into their quark-antiquark constituents. In the instanton vacuum,
quarks acquire an effective mass as a consequence of the spontaneous
breaking of chiral symmetry. Such a "constituent" quark mass at rest
$M_q$ was calculated in several approaches and  found to range from
$\simeq~400~$MeV to $200~$ MeV, depending on the bare quark mass
(see e.g.\cite{musakh} and references therein). In
Fig.~\ref{fig:rho}, we compare the effective mass plot for the
$\rho$ meson with the $2~M_\pi$ line, and the $2 M_q$ line
calculated in our model. We can see that, for the two largest quark
masses, the effective mass falls below the $2 M_\pi$ line, but
always remains above the $2 M_q$ line. Hence, there is always phase
space available to decay into constituent quarks.

\section{Conclusions}
\label{conclusions}

In this work we have studied the contribution of  instanton-induced
chiral forces in the $a_1$ and $\rho$ meson resonances.

We have provided clean evidence that the $a_1$ and $\rho$ meson  can
exist even in the presence of instanton forces only (i.e. in the
instanton vacuum). Their masses extracted from a fit of the
effective mass plot have been found to be about $30\%$ larger than
the corresponding experimental values. This deviation should be
confronted with the excellent agreement with the experimental masses
previously found in the case of the pion and the nucleon
\cite{IILMchpt}. These results suggest that chiral forces are weaker
in these resonances, but still represent the leading source of
interactions.

On the other hand, the confining forces are very important to
determine  the width of these resonances and their decay properties.
In fact, we have observed that, in this model, the vector meson is
unstable even when quark masses are large and there is no phase
space to decay into two pions. This can be explained assuming that
--- as a consequence of the absence of confinement--- mesons can
dissociate into their constituents.

In the future, it would be interesting to study how the stability of
the meson  is restored, once confinement is introduced in the model.
A way to do so would be to extend the ensemble of gauge
configurations to include regular-gauge instantons and to study the
behavior of the effective mass as a function of the density of such
pseudo-particles~\cite{reginst}.


\acknowledgements We thank John W. Negele for help and illuminating
discussions. This work was supported in part by the U.S. DOE office
of Nuclear Physics under contract DE-FC02-94ER40818 and the INFN-MIT
"Bruno Rossi" Exchange Program.
\bibliography{ILM}
\end{document}